\documentclass[pre,a4paper,twocolumn]{revtex4}
\usepackage{graphicx}
\usepackage{amsmath}
\usepackage{psfrag}
\usepackage{latexsym}

\newcommand{\bq}{\begin{equation}}
\newcommand{\eq}{\end{equation}}
\newcommand{\ba}{\begin{eqnarray}}
\newcommand{\ea}{\end{eqnarray}}
\newcommand{\dd}{{\rm d}}

\begin{document}
\title{Condensation in zero-range processes on inhomogeneous networks}

\author{B. Waclaw$^{1,3}$}\thanks{bwaclaw@th.if.uj.edu.pl}
\author{L. Bogacz$^{1,3}$}{\thanks{bogacz@th.if.uj.edu.pl}
\author{Z. Burda$^{1,2}$}{\thanks{burda@th.if.uj.edu.pl}
\author{W. Janke$^{3,4}$}{\thanks{Wolfhard.Janke@itp.uni-leipzig.de}
\affiliation{$^1$Marian Smoluchowski Institute of Physics,
Jagellonian University, Reymonta 4, 30-059 Krak\'ow, Poland \\
$^2$Mark Kac Complex Systems Research Centre, Jagellonian University, Krak\'ow,
Poland \\
$^3$Institut f\"ur Theoretische Physik, Universit\"at Leipzig,
Augustusplatz 10/11, 04109 Leipzig, Germany \\
$^4$Centre for Theoretical Sciences (NTZ), Universit\"at Leipzig, Germany}

\begin{abstract}
We investigate the role of inhomogeneities in zero-range processes in
 condensation dynamics.
We consider the dynamics of balls hopping 
between nodes of a network, and find that the condensation 
is triggered by the ratio $k_1/k$ of the highest degree $k_1$ 
to the average degree $k$. Although the condensate takes 
on the average an extensive number of balls, its occupation 
can oscillate in a wide range. We show that in systems 
with strong inhomogeneity, the typical melting time 
of the condensate grows exponentially with the 
number of balls.
\end{abstract}
\maketitle

\section{Introduction}
Zero-range processes \cite{cc1,cc2,cc3,cc4,cc5,cc6} have attracted attention of many researchers
since they provide an exactly solvable example of 
far-from-equilibrium dynamics and of condensate formation.
Many questions concerning the dynamics of the model
can be addressed and solved analytically. It is known that a  
zero-range process has a steady state and that this state 
is described by the partition function of the balls-in-boxes (B-in-B) 
model \cite{bbj,bbj2}, also called the Backgammon model. 
The B-in-B model has two phases: a fluid and a condensed 
one, separated by a critical point at which the system 
undergoes a phase transition and the condensate is formed. 
Unlike the Bose-Einstein condensation,
the B-in-B condensation takes place in real space
rather than in momentum space. 
Therefore it mimics such processes like mass transport \cite{cc1}, 
condensation of links in complex networks \cite{cc2,bbw} or phase separation \cite{ph1,ph2}.

The zero-range process (ZRP) \cite{evans} discussed in this paper describes
a gas of identical, indistinguishable 
particles hopping between the neighboring nodes of a network.
The state of such a system is characterized by the topology of 
the network, which is fixed during the process, and
by the particle distribution which is given by
the occupation numbers of particles $\{m_i\}$ on all nodes 
$i=1,\ldots, N$ of the network.
The total number of particles $M=m_1+m_2+\ldots + m_N$ 
is conserved during the process. 
The zero-range dynamics is characterized by the outflow rates $u(m)$ of particles from network nodes,
which depend only on the occupation number $m$ of the node from which the particle hops.
We shall assume that these ultra local hopping rates
are identical for each node. The stream of particles 
outgoing from a node is equally  distributed among all its
neighbors. So if we denote by $k_i$ the number of neighbors of the node $i$, called also its degree, 
then the hopping rate from the $i$-th node to each of its neighbors is equal to $\frac{1}{k_i} u(m_i)$.
The outflow rate $u(m)$ is a semi-positive function which is 
equal to zero for $m=0$. For a given graph the function
$u(m)$ entirely defines the dynamics of the system. 

We consider here the ZRP on a network being a connected simple
graph. In this case, the ZRP has a unique steady state,
in which the probability $P(m_1,\dots,m_N)$ 
of finding the distribution of balls $\{m_1,\ldots,m_N\}$ 
factorizes into a product of some weight functions
$p_i(m_i)$ for individual nodes, except that there is a global constraint reflecting the conservation of particles.  

On a $k$-regular network, that is when all node degrees are equal to $k$, 
all weight functions $p_i(m)$ are identical, $p_i(m)\equiv p(m)$, 
and thus the probability $P$ is invariant under permutations 
of the occupation numbers. When the density
$\rho=M/N$ of balls per node exceeds a certain critical value $\rho_c$ depending on the functional form of $p(m)$, 
a single node attracts an extensive number of balls called the condensate. The relative occupation of that node does not disappear
in the thermodynamic limit $N,M\to\infty$, with fixed $\rho$. The larger the density $\rho$, the larger is the
number of balls in the condensate. In other words,
the system undergoes a phase transition at $\rho=\rho_c$ 
between the fluid (low density) and the condensed (high density)
phase. The permutation symmetry of $\{m_i\}$ is respected in the
fluid phase while it is broken in the condensed phase where one node
becomes evidently distinct from the $N\!-\!1$ remaining ones.
The symmetry of the partition function reduces to the subgroup of permutations
of $N\!-\!1$ occupation numbers. This mechanism 
has been extensively studied in the B-in-B model \cite{bbj,bbj2,bbjw}.
The value of the critical
density depends on the weight function
$p(m)$ which can be translated to the asymptotic properties of $u(m)$ for $m\to\infty$.
If $u(m)$ tends to infinity
then also $\rho_c=\infty$ and the condensation does not occur regardless
of the density of balls. The system is
in the fluid phase for any finite density $\rho$. Intuitively,
this means that there exists an effective repulsive force 
preventing a node from being occupied by many balls
and they distribute uniformly on the whole graph.
On the contrary, if $u(m) \to 0$, 
the critical density is $\rho_c=0$ and therefore the system is in the condensed phase
for any $\rho>0$. The larger the number of particles on a node, the smaller
are the chances for balls to escape from it since $u(m)$ becomes very small
for large $m$. This can be seen as the existence of an effective attraction between particles.

The most interesting case is when $u(m)$ goes to some positive constant $u_\infty$ with $m\to\infty$. 
One can show that the probability distribution
$P(m_1,\ldots,m_N)$ does not depend on $u_\infty$ \cite{bbjw}
but on how fast $u(m)$ approaches the constant value.
Therefore without loss of generality we can choose $u_\infty=1$ and concentrate on the asymptotic behavior
having the form: $u(m)=1+b/m$, with $b$ being some positive number.
If $0\le b \le 2$, then the critical density $\rho_c$ is infinite.
The effective attraction between balls is too weak to form the condensate.
However, if $b>2$, then $\rho_c$ has a finite value.  
In this case the attraction is strong enough to trigger 
the condensation above the critical density $\rho_c$. 

So far we have discussed the criteria of the condensation on $k$-regular networks, where it arises
as a result of the spontaneous symmetry breaking. All those facts are well known \cite{evans,evans2}.
On the other hand, the permutation symmetry can 
be explicitly broken if the weight functions $p_i(m)$  
are not identical for all nodes. For instance, this happens when the network 
on which the process takes place is inhomogeneous, that is when
the degrees $k_1,\dots,k_N$ vary. A particularly important example are complex networks \cite{cn},
for which the distribution of degrees has usually a long tail, and thus there are
many nodes with relatively high degree. They are, however, not easy for analytical studies although 
some predictions are possible \cite{jdn}.
Below we shall argue that to gain some insight into the static and dynamical properties
of the ZRP on such networks it is sufficient to study some simplified models.

In the remaining part of the paper we 
consider only the most favorable situation when
the weight of only one node is different from 
the remaining ones. Such an inhomogeneity of the weights
can be introduced either by an inhomogeneity of the outflow
rates $u_i(m)$ or by an inhomogeneity of the degree distribution.
We focus here on the latter situation, when a graph has one node of 
degree $k_1$ which differs from all
remaining degrees $k_2=\ldots =k_N\equiv k$. As we shall see,
in this case the quantity $\mu = \log(k_1/k)$ plays the role
of an external field breaking the permutation symmetry.

In particular, we discuss the dynamics of the condensate on such inhomogeneous networks.
This is a relatively new topic and, in contrast to the stationary properties, less understood.
Although the emergence of the condensate has been studied for homogeneous and inhomogeneous systems \cite{evans} and numerically for scale-free networks in \cite{jdn}, studies of the dynamics of an existing condensate
are rare \cite{god}.
For instance, one question which may be asked is what is the typical life time of the condensate,
that is how much time does it take to ``melt'' the condensate at one node and rebuild it at another node.
To provide an answer to this problem is the main goal of the present article.

The rest of the paper is organized as follows. In Sec.~II we 
discuss the static properties of the ZRP on inhomogeneous networks. 
We consider some particular graph topologies: $k$-regular graphs,
star graphs and $k$-regular graphs with a single inhomogeneity
introduced by a vertex of degree $k_1>k$. Their advantage is that
all calculations can be done exactly or at least with an excellent approximation.
In all cases we calculate the effective occupation number distribution
$\pi(m)$ and use it to derive information about the condensate
dynamics. In Sec.~III  we derive
analytic expressions for the life time of the condensate.
We concentrate on the role of
inhomogeneity and typical scales at which it becomes relevant.
All analytical results are cross-checked by Monte Carlo simulations.
The last section is devoted to a summary of our results.

\section{Steady state -- statics}
A zero-range process on a connected simple graph has a steady state
with the following partition function $Z(N,M,\{k_i\})$ 
\cite{evans}:
\bq
Z(N,M,\{k_i\}) = 
\sum_{m_1=0}^M \cdots \sum_{m_N=0}^M \delta_{ \sum_{i=1}^N m_i, M}
\prod_{i=1}^N p(m_i) k_i^{m_i},	\label{part}
\eq
where $\delta_{i,j}$ denotes the discrete delta function and the weight function $p(m)$ is related to the hop rate $u(m)$ through the formula:
\bq
p(m)= \prod_{n=1}^m \frac{1}{u(n)}, \;\;\; p(0)=1.	
\label{pbyu}
\eq
We shall denote $Z(N,M,\{k_i\})$ in short by $Z(N,M)$, having in mind its dependence on the degrees. 
The partition function (\ref{part}) contains the whole 
information about the static properties of the system in the steady state.
The only trace of graph topology in the formula is through
the nodes degrees. The dynamics, however, depends also on other topological 
characteristics but they become important only in refined treatments.
In a sense, the degree sequence is the first-order approximation also for the dynamics. 

The probability $P(m_1,\dots,m_N)$ of a given configuration 
$\{m_i\}$ reads:
\ba
P(m_1,\dots,m_N) &=& \frac{1}{Z(N,M)} 
\prod_{i=1}^N p(m_i) k_i^{m_i} \nonumber \\
&=& \frac{1}{Z(N,M)} \prod_{i=1}^N \tilde{p}_i(m_i),
\ea
where we have defined ``renormalized'' weights
$\tilde{p}_i(m_i) = p(m_i)k_i^{m_i}$, being now node-dependent.
The most important quantity characterizing the steady state is
the probability $\pi_i(m)$ that the
$i$-th node is occupied by $m$ particles:
\ba
\pi_i(m_i) = \sum_{m_1} \cdots \sum_{m_{i-1}} 
\sum_{m_{i+1}} \cdots \sum_{m_N} P(m_1,\dots,m_N) \times \nonumber \\
\times \delta_{\sum_{j=1}^N m_j, M} = \frac{Z_i(N-1,M-m_i)}{Z(N,M)} \tilde{p}_i(m_i),
\label{pigeneral}
\ea
where $Z_i(N-1,M-m)$ denotes the partition function 
for $M-m$ particles occupying a graph consisting 
of $N-1$ nodes with degrees $\{k_1,\dots,k_{i-1},k_{i+1},\dots,k_N\}$. 
We shall call $\tilde{p}_i(m)$ ``bare'' occupation probability
while $\pi_i(m)$  ``dressed'' or effective occupation probability
of the node $i$. We also define the average occupation probability:
\bq
\pi (m) = (1/N) \sum_i \pi_i(m).
\eq
For a $k$-regular graph, that is for $k_i\equiv k$,
the occupation probability $\pi_i(m)=\pi(m)$ is the same for every
node and all the formulas above reduce to those discussed 
in \cite{bbj}. In general, the partition function can be calculated recursively:
\ba
& & Z(N,M,\{k_1,\dots,k_N\}) = \nonumber \\
&=& \sum_{m_N} \tilde{p}_N(m_N) \sum_{m_1,\dots,m_{N-1}} \delta_{\sum_{i=1}^{N-1} m_i, M-m_N} \prod_{i=1}^{N-1} \tilde{p}_i(m_i) \nonumber \\
& =& \sum_{m_N=0}^{M} \tilde{p}_N(m_N) Z(N-1,M-m_N,\{k_1,\dots,k_{N-1}\}). \nonumber \\
\label{znmrec}
\ea
For $N=1$ the partition function simply reads 
$Z(1,M,k_1) = \tilde{p}_1(M)$. The recursive use of the
formula (\ref{znmrec}) allows one to compute 
the partition function within a given numerical accuracy.
Using this method we were able to push the computation as far as to
$N$ of order $500$. For identical weights, one can find
a more efficient recursion relation by splitting the system into
two having similar size, which allows to study much larger systems.
The computation of the partition function can be used together
with Eq.~(\ref{pigeneral}) to determine 
numerically the node occupation distribution $\pi_i(m)$. 
This gives an exact result with a given accuracy and is more efficient than the corresponding 
Monte Carlo simulations of the ZRP. The dynamics, however, is not accessible in this way.

As mentioned in the introduction we are going to examine the effect of
topological inhomogeneity on the properties of the ZRP.
We shall consider an almost $k$-regular graph 
with one node, say number one, having a degree bigger than the rest of nodes:
$k_1>k$ and $k_2=\dots=k_N=k$. 
The simplest realization of such a graph is a star or a wheel graph.
In general, such a graph can be constructed from any $k$-regular
graph by a local modification. We proceed as follows. First we
 derive exact formulas for the particle distribution
in a steady state on a $k$-regular graph which shall later
serve us as a reference point. Then we consider the particular
example of a star topology as the simplest example
of a single defect and finally an arbitrary $k$-regular graph with a singular node $k_1>k$.
The system has now the following weights: 
$\tilde{p}_1(m)= k_1^m p(m)$ for the singular and $\tilde{p}_i(m)= k^m p(m)$ for the regular nodes. 
They differ by an exponential factor
$\tilde{p}_1(m)/\tilde{p}_i(m)= (k_1/k)^m = e^{\mu m}$
where $\mu =\log(k_1/k)>0$, which clearly favors the situation
in which the singular node has much more particles
than the regular ones. To make things as simple as possible, 
and to concentrate on the effect of inhomogeneity
we assume that the outflow rate $u(m)=1$ is constant
and independent of $m$. In this case $p(m)=1$ is also constant,
simplifying calculations.
All other functions with the asymptotic behavior 
$u(m)\rightarrow 1$ would lead basically to the same qualitative 
behavior. This is because
in this case $p(m)$ would have a power-law tail which is much
less important for the large $m$-behavior than the exponential
factor $e^{\mu m}$ introduced by the inhomogeneity.
We shall briefly comment on this towards the end of the paper.

\subsection{$k$-regular graph}
With the assumption $p(m)=1$,
the partition function $Z(N,M)$ from Eq.~(\ref{part}) for the steady state of the ZRP on a $k$-regular graph reads:
\ba
Z_{\rm reg}(N,M) &=& \sum_{m_1=0}^M \cdots \sum_{m_N=0}^M 
\delta_{\sum_{i} m_i, M} \;
k^{\sum_i m_i} \nonumber \\
&=& k^M \frac{1}{2\pi i} \oint \dd z \, 
z^{-M-1} \left(\sum_{m=0}^M z^m\right)^N.
\label{cint}
\ea
We used an integral representation of the discrete delta function
which allowed us to decouple the sums over $m_1,\dots, m_N$ for
the price of having the integration over $z$.
The sum over $m$ can be done yielding $1/(1-z)$. Using
the expansion:
\bq
\left(\frac{1}{1-z}\right)^N = 
\sum_{m=0}^\infty \binom{-N}{m} (-z)^m 
= \sum_{m=0}^\infty \binom{N+m-1}{m} z^m
\eq
and Cauchy's theorem we see that the contour
integration over $z$ selects only the term with $m=M$
from the integrand in (\ref{cint}), so we obtain:
\bq
Z_{\rm reg}(N,M) = k^M \binom{N+M-1}{M}. \label{zreg}
\eq
Inserting this into Eq.~(\ref{pigeneral})
we find the occupation number distribution
\ba
\pi(m) &=& \pi_i(m) = \binom{M+N-m-2}{M-m} / \binom{M+N-1}{M}  \nonumber \\
&\propto & \frac{(M+N-m-2)!}{(M-m)!}. \label{pik}
\ea
The distribution $\pi(m)$ is identical for all nodes and independent on $k$. 
It falls faster than exponentially for large $m$, therefore the condensate
never appears.

In particular one can apply these formulas to the complete 
graph which is just a $k$-regular graph with $k=N-1$.
In Fig.~\ref{f1} we see the comparison between the
theoretical expression (\ref{pik}) and results of
numerical Monte Carlo simulations for a $4$-regular graph with $N=20$ nodes and 
two different numbers of balls, $M$. 
The simulations of the
ZRP were organized in sweeps consisting on $N$ steps each. In a single
step a node was chosen at random and if it was non-empty
a particle was picked and moved to a neighboring node.
For each graph the process was initiated from a random
distribution of particles. After some thermalization,
 measurements of $\pi(m)$
were done on $10^4$ configurations generated every sweep.
The results were averaged over these configurations and then
over $5\times 10^4$ independent graphs drawn at random
from the ensemble of $k$-regular graphs. 

\begin{figure}
\psfrag{x}{$m$}
\psfrag{y}{$\pi(m)$}
\includegraphics[width=8cm]{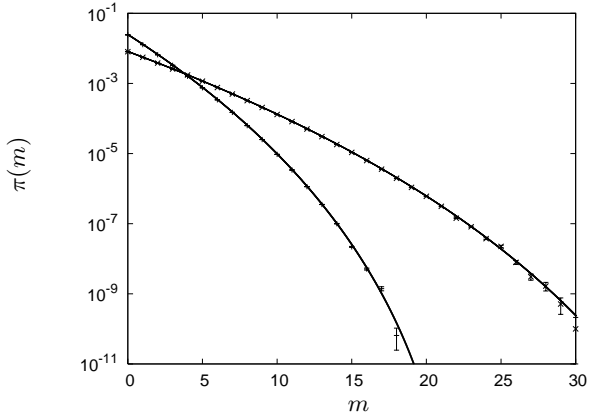}
\caption{The ``experimental'' distribution $\pi(m)$ compared to 
the theoretical prediction (\ref{pik}) for regular graphs 
with $k=4$ and $N=20$ and for $M=20$ (+) and $M=40$ (x) balls.}
\label{f1}
\end{figure}

\subsection{Star graph}
We consider first a special case of a single 
inhomogeneity graph, namely the star graph having $N-1$ nodes of 
degree $k_2=\ldots=k_N=1$ connected to the central node with $k_1=N-1$.
The partition function $Z_{\rm star}(N,M)$ is
\ba
Z_{\rm star}(N,M) &=& 
\sum_{m_1=0}^\infty \cdots \sum_{m_N=0}^\infty \delta_{\sum_{i=1}^N m_i , M}
(N-1)^{m_1} \hspace{5mm} \nonumber \\
&=& \sum_{m=0}^M (N-1)^m \binom{M+N-m-2}{M-m}, \label{zstar}
\ea
as follows from Eq.~(\ref{zreg}). It is convenient to change
the summation index from $m$ to $j = M-m$ which can be interpreted as
a deficit of particles counted relatively to the full occupation:
\ba
Z_{\rm star}(N,M) = (N-1)^M \sum_{j=0}^M (N-1)^{-j} \binom{N+j-2}{j} . \nonumber \\
\ea
Let us assume that $N\gg 1$. The summands in the last expression are strongly
suppressed when $j$ increases so the sum can 
be approximated by changing the upper limit from $M$ to $\infty$. We obtain
\ba
Z_{\rm star}(N,M)  
&\cong & (N-1)^M \sum_{j=0}^\infty \left(\frac{-1}{1-N}\right)^j 
\binom{-(N-1)}{j} \nonumber \\
&=& (N-1)^M \left(1-\frac{1}{N-1}\right)^{1-N} \nonumber \\
&=& (N-1)^M \left( \frac{N-1}{N-2} \right)^{N-1}.
\ea
Using Eq.~(\ref{pigeneral}) 
and the partition function $Z_{\rm reg}$ 
calculated in Eq. (\ref{zreg}) of the previous subsection we can determine the distribution of particles
$m$ at the central (singular) node,
\ba
&\pi_1(m) &= \frac{Z_{\rm reg}(N-1,M-m)}{Z_{\rm star}(N,M)} (N-1)^m \nonumber \\
&=& (N-1)^{m-M} \binom{M+N-m-2}{M-m} \left( \frac{N-2}{N-1} \right)^{N-1}. \nonumber \\
\label{pi1star}
\ea
Similarly, we can determine the distribution of particles
on any external (regular) node $i$:
\bq
\pi_{i}(m) = \frac{(N-2)(N-1)^{-m}}{N-1-(N-1)^{-M}} \approx \frac{N-2}{(N-1)^{m+1}}. \label{pistext}
\eq
We see that $\pi_{i}(m)$ decays exponentially 
with $m$ while $\pi_1(m)$ grows exponentially for $m\ll M$: 
\bq
\pi_1(m) \propto e^{ m\left( 
-\frac{1}{2M}+\frac{1}{2(M+N-2)} +\log\frac{M(N-1)}{M+N-2} \right)}.
\eq
The growth slows down for $m$ approaching $M$. 
At $m=M$, $\pi_1(m)$ reaches its maximal value:
\bq
\pi_1(M) = \left(\frac{N-2}{N-1}\right)^{N-1},
\eq
which tends to $e^{-1}$ when $N$ goes to infinity.
In Fig.~\ref{f2} we compare the theoretical distributions 
(\ref{pi1star}) and (\ref{pistext}) with Monte Carlo 
simulations of the ZRP for $N=20$ nodes and $M=20, 30, 40$. 
The agreement is very good.
\begin{figure}
\psfrag{x}{$m$}
\psfrag{y}{$\pi(m)$}
\includegraphics[width=8cm]{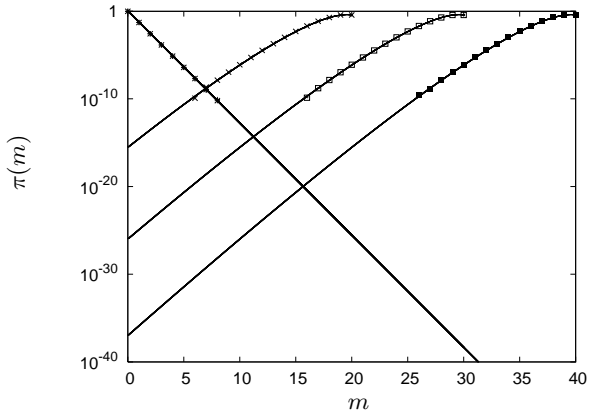}
\caption{The ``experimental'' and the theoretical (solid lines) 
particle number distributions for the star graph, 
for $N=20$, and $M=20$ (crosses), $30$ (empty squares) and $40$ (filled squares). 
The theoretical distributions were calculated according to the
formula (\ref{pi1star}) for the central node (rising curves) 
and according to Eq.~(\ref{pistext}) for external nodes (falling curve, the Monte Carlo data (points) plotted for $N=20$ and $M=30$). 
}
\label{f2}
\end{figure}

It is instructive to calculate the mean number 
of particles at the central node. To simplify calculations
we make use of the distribution $\widehat{\pi}_1(j)\equiv \pi_1(M-j)$ of
the deficit of particles defined above:
\bq
\widehat{\pi}_1(j) = 
(1-\alpha)^{1-N} (-\alpha)^j \binom{-(N-1)}{j}.
\label{remain}
\eq
The parameter $\alpha=1/(N-1)$ is just the ratio of any external node degree to the degree of the central node and measures
the level of inhomogeneity. 
The overall prefactor $(1-\alpha)^{1-N}$ is independent of $j$. It is just a normalization constant
that results from summing the $j$-dependent part of the expressions (\ref{remain}),
\bq
S(\alpha) = \sum_{j=0}^\infty 
(-\alpha)^j \binom{-(N-1)}{j} = 
\frac{1}{(1-\alpha)^{N-1}} .
\label{auxs}
\eq
As before we changed the upper limit from
$M$ to infinity because $\alpha \ll 1$ for the star graph and hence
the summands are strongly suppressed for large $j$.
The average deficit at the central node is
\bq
\left<j_1\right> = 
\sum_{j=0}^M \widehat{\pi}_1(j) j = 
\alpha \frac{\dd \log S(\alpha)}{\dd\alpha} = \frac{N-1}{N-2}, 
\label{m1star}
\eq
as follows from Eq.~(\ref{auxs}). For large $N$ it tends to 
one, so we have
\bq
\left<m_1\right> = M - \left<j_1\right> \cong M-1.
\eq
We see that on average almost all balls are concentrated at the central
node and only one ball is in the rest of the system.
We can also determine the range of fluctuations around $\left<m_1\right>$
by calculating the variance. Taking again advantage of the generating function (\ref{auxs}) one finds
\ba
\left<(m_1-\left<m_1\right>)^2\right> = \left<(j_1-\left<j_1\right>)^2\right>  \nonumber \\
= \frac{\dd^2 \log S(e^{-\mu})}{\dd\mu^2} = \left(\frac{N-1}{N-2}\right)^2,
\ea
where we used the parameter $\mu=-\log\alpha=\log (N-1)$.
For large $N$ the result tends to one, so
we can draw the following picture. 
For any $N\gg 1$ we observe a condensation of 
particles at the central node regardless of their density $\rho=M/N$. 
The critical density is equal to zero and the system is always in the condensed phase. 
The condensate residing at the central node contains $M-1$ particles, with very small fluctuations,
while the other nodes are almost empty. 

\subsection{Single inhomogeneity}
A very particular property of the star graph is that the
inhomogeneity increases with its size.
Therefore, it is interesting to consider the situation
when the inhomogeneity $\alpha = k/k_1$ is arbitrary and independent of $N$. 
The single inhomogeneity graph we consider here has one node of degree 
$k_1$ and $N-1$ nodes of degree $k$. Again, we assume that $k_1>k$.
The partition function (\ref{part}) takes now the form
\bq
Z_{\rm inh}(N,M) = 
\sum_{m_1=0}^M k_1^{m_1} 
\sum_{m_2,\dots,m_N=0}^M \delta_{\sum_i m_i,M} 
\; k^{\sum_{i=2}^N m_i}.
\label{zsh1}
\eq
The sum over $m_2,\dots,m_N$ is equal to the partition function 
$Z_{\rm reg}(N-1,M-m_1)$ given by Eq.~(\ref{zreg}).
The whole formula looks almost identical to that
for the star graph except that
now the degree $k_1$ does not need to be much greater than 
$k$ and therefore the substitution of $M$
by $\infty$ has to be done carefully in a manner
incorporating finite-size corrections. As before, we first
change variables from $m_1$ to $j = M-m_1$. Using
Eq.~(\ref{auxs}) we can cast the formula (\ref{zsh1}) into the 
following form:
\ba
Z_{\rm inh}(N,M) &=& 
k_1^M \sum_{j=0}^M \alpha^j 
\binom{N+j-2}{j} \nonumber \\
&=& k_1^M \left[ (1-\alpha)^{1-N} - c(M) \right].
\label{zinh}
\ea
The correction $c(M)$ is equal to the sum over $j$ from $M+1$ to infinity.
It corresponds to the surplus which has to be
subtracted from the infinite sum represented by the first
term in square brackets. In the limit $M\to\infty$ it can be estimated
as follows:
\bq
c(M) = \sum_{j=M+1}^\infty \alpha^j \binom{N+j-2}{j} 
\approx \frac{1}{(N-2)!} \int_M^\infty \dd j \, e^{F(j)},
\label{intc}
\eq
where 
\bq
F(j) = j \log\alpha + \log\left((N+j-2)!\right) - \log(j!).
\eq
Using Stirling's formula we can calculate the 
integral (\ref{intc}) by the saddle-point method.
Taking into account only leading terms we have
\ba
& &\int_M^\infty \dd j e^{F(j)}  \approx e^{F(j_*)} \times \nonumber \\
& &\times \sqrt{\frac{-\pi}{2F''(j_*)}} \mbox{erfc}\left( (M-j_*)
\sqrt{-F''(j_*)} \right),
\ea
where erfc denotes the complementary error function,
\bq
\mbox{erfc}(x) = \frac{2}{\sqrt{\pi}} \int_x^\infty \dd y e^{-y^2} ,
\eq
and  $j_*$ is determined from the saddle equation $F'(j_*)=0$:
\bq
j_* \approx \frac{\alpha (N-2)}{1-\alpha}.
\eq
Collecting all terms we eventually find
\ba
c(M) &\approx &
\frac{\alpha^{\frac{\alpha(N-2)}{1-\alpha}}}{1-\alpha} \frac{((N-2)/(1-\alpha))!}
{(\alpha(N-2)/(1-\alpha))!}\sqrt{\frac{\pi \alpha (N-2)}{2}} \times \nonumber \\
&\times & \frac{1}{(N-2)!} \mbox{erfc}\left( \frac{M(1-\alpha)-\alpha(N-2)}{\sqrt{\alpha(N-2)}}\right)	. 
\label{cmfin}
\ea
In order to keep formulas shorter we used here the notation
$x! \equiv \Gamma(1+x)$ also for non-integer arguments.
The complete partition function 
$Z_{\rm inh}(N,M)$ is given by the right-hand side of Eq.~(\ref{zinh}) 
with $c(M)$ given by Eq.~(\ref{cmfin}). 
We can now calculate $\pi_1(m)$, that is the distribution
of balls at the singular node,
\bq
\pi_1(m) = \frac{Z_{\rm reg}(N-1,M-m)}{Z_{\rm inh}(N,M)} k_1^m,
\eq
where $Z_{\rm reg}$ is the partition function (\ref{zreg}) for a 
regular graph with degree $k$. Using Eqs.~(\ref{zreg}) and (\ref{zinh})
we obtain
\bq
\pi_1(m) = \binom{M+N-m-2}{M-m} \frac{\alpha^{M-m}}{(1-\alpha)^{1-N} - c(M)}. \label{pi1inh}
\eq
In Fig.~\ref{f3} we show the theoretical ball distributions for graphs 
with $k=4$, $N=20$ and various $M$ for singular nodes with $k_1=8$ and $k_1=16$, respectively,
and compare them with the corresponding results obtained
by Monte Carlo simulations. The agreement, although very good for the presented plots, is the better, the smaller is the ratio $\alpha=k/k_1$.
\begin{figure}
\psfrag{x}{$m$}
\psfrag{y}{$\pi(m)$}
\psfrag{a}{$\alpha=\frac{1}{2}$}
\psfrag{a2}{$\alpha=\frac{1}{4}$}
\includegraphics[width=8cm]{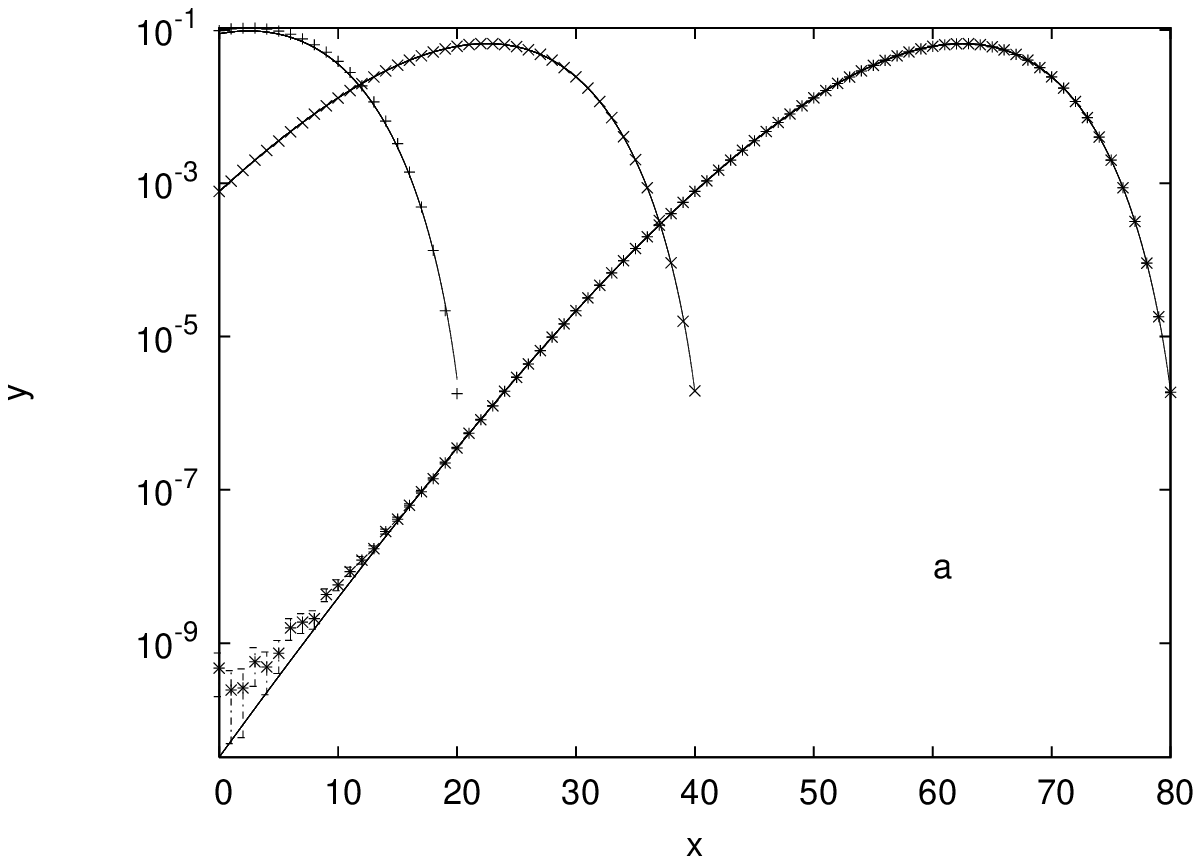}
\includegraphics[width=8cm]{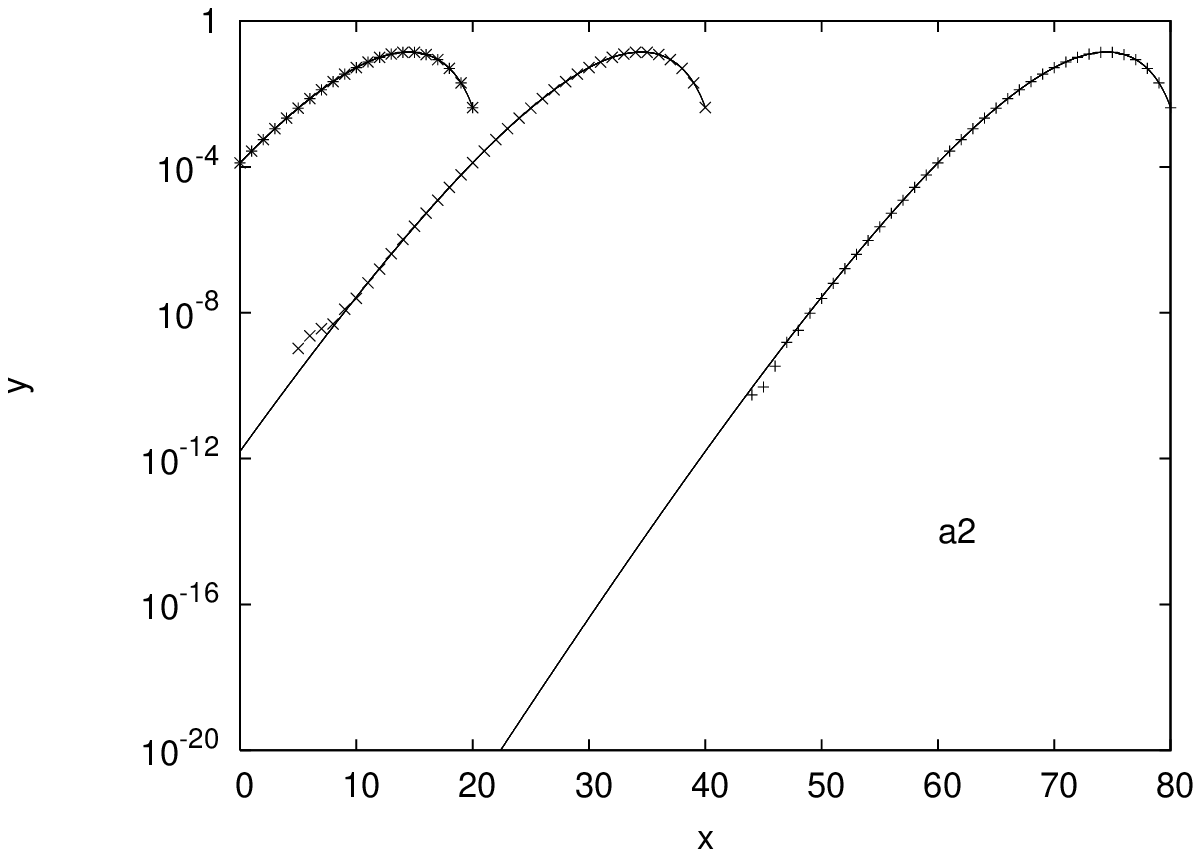}
\caption{The distribution of balls at the singular node for graphs 
with $k=4$, $N=20$, $k_1=8$ (top) and $k_1=16$ (bottom). 
The total number of balls is $M=20,40$ and $80$ from left to 
right curve, respectively. Points represent numerical data while solid 
lines show Eq.~(\ref{pi1inh}).
}
\label{f3}
\end{figure}
Neglecting an inessential normalization, we see that
Eq.~(\ref{pi1inh}) has the asymptotic behavior
\bq
\pi_1(m) \propto \left(\frac{k_1}{k}\right)^m 
\binom{M+N-m-2}{M-m} \sim \exp(G(m)),
\eq
where 
\ba
G(m) &=& \left(M+N-m-\frac{3}{2}\right)\log(M+N-m-2)- \nonumber \\
&-& m \log\alpha - \left(M-m+\frac{1}{2}\right) \log(M-m).
\label{gfunc}
\ea
The number of particles of the condensate can be 
estimated using the saddle point equation
$G'(m_*)=0$ for $m_*>0$. Neglecting terms 
of order $1/M^2$ we find
\bq
m_* \cong M - \frac{\alpha}{1-\alpha} (N-2).
\eq
Alternatively one can calculate the number of particles
of the condensate as the mean of the distribution $\pi_1(m)$.
Adapting the same trick as in the previous section,
\ba
\left<m_1\right> &=& M - \left<j_1\right> = 
M - \alpha \frac{\dd \log S(\alpha)}{\dd\alpha} \nonumber \\
&=& M - \frac{\alpha}{1-\alpha} (N-1) \approx m_*,
\label{m1inh}
\ea
as follows from Eqs.~(\ref{auxs}) and (\ref{m1star}). The criterion for
condensation is that the central node contains an extensive
number of balls. In the limit $N,M\to \infty$ and fixed density
$\rho=M/N$ it amounts to the condition $\left<m_1\right>>0$ 
leading to the critical density
\bq
\rho_c = \frac{\alpha}{1-\alpha}.
\eq
The condensation takes place when $\rho>\rho_c$ exactly like 
in the Single Defect Site model \cite{evans}. 
The critical density decreases with decreasing ratio $\alpha=k/k_1$ or, equivalently, 
with increasing ``external field'' $\mu=\log(k_1/k)$. The singular node attracts
$N(\rho-\rho_c)+\rho_c$ balls on average as 
follows from Eq.~(\ref{m1inh}).
It is also easy to find that the distribution of balls 
$\pi_i(m)$ at any regular node falls exponentially,
\bq
\pi_i(m) \sim \left(\frac{k}{k_1}\right)^m = \alpha^m = e^{-\mu m},
\eq
thus the condensate never appears on it. 
A regular node contains on average $\left<m_i\right> = \rho_c$ balls independently of the total density $\rho$ of balls in the system
as long as it exceeds $\rho_c$.

\section{Dynamics of the condensate}

Let us now turn to a discussion of the dynamics of the condensate.
From the previous section we know that the condensate
spends almost all time at the node with highest degree. However,
 occasionally it ``melts'' and disappears from the singular node for a short while.
We know that the probability of such an event is very small, so we expect 
the life-time of the condensate to be very large. 
Following the ideas of \cite{god}, let us imagine that we monitor only
the number of particles at the singular node, which fluctuates in time. 
The temporal sequence of occupation numbers at this node
performs a sort of one-dimensional random walk 
and can be viewed as a Markov chain. Using a mean-field approximation
one can derive effective detailed balance equations for 
the incoming and the outgoing flow of particles for this node.
The approximation is based on the assumption that the
remaining part of the system is quickly thermalized, much faster
than the typical time scale of the melting process on the monitored node.
Thus the balance equations are written for the singular node and a single mean-field node having 
some typical properties. 
For this mean-field dynamics one can derive many quantities of interest.
In particular it is convenient to calculate the average 
time $\tau_{mn}$ it takes to decrease the occupation number of
the monitored node from $m$ to $n$,
or more precisely, the average first passage time for the Markov process
initiated at $m$ to pass $n$. This quantity was first derived 
in \cite{god} for the ZRP on a complete graph with outflow rates 
$u(m) = 1+b/m$. The formula derived there,
\bq
\tau_{mn} = \sum_{p=n+1}^{m} \frac{1}{u(p)\pi(p)} \sum_{l=p}^{M} \pi(l),
\label{Tmngen}
\eq
can be easily adapted to the case discussed in our paper by 
setting $u(p)=1$ and using the distribution $\pi_1(p)$ of the singular node
in place of $\pi(p)$ in the original formula.
Equipped with the formula for $\tau_{mn}$ we are in principle able to calculate the 
typical melting time $\tau$. What is yet missing is the condition for $n$ at
which the condensate can be considered as completely melted. 
We shall choose the simplest possible criterion and define the
``typical'' melting time $\tau$ as $\tau_{m0}$, that is the time needed to completely empty 
the monitored node beginning from $m$ equal to the average occupation 
of the node in the steady state. 

It was shown \cite{god} that for the complete graph and $u(m) = 1+b/m$, the melting time
is approximately given by
\bq
\tau\propto (\rho-\rho_c)^{b+1} M^b,
\eq
where $\rho_c$ is the critical density above which the condensate
is formed. The power-law increase with $M$ can be attributed
to the power-law fall of $\pi(m)\sim m^{-b}$, characteristic for homogeneous systems
with $u(m)=1+b/m$.
The key point of our paper is that for inhomogeneous networks the melting time does no longer 
follow a power-law but instead increases {\em exponentially} with $M$ due to the occurrence
of the inhomogeneity which can be regarded as an external field $\mu = \log (k_1/k)$, breaking the symmetry.

Before we do the calculations let
us make a general remark about the dependence of $\tau_{mn}$ on $m$ and $n$. A quick inspection of Eq.~(\ref{Tmngen}) 
tells us that significant contribution to the sum over $p$ comes from terms  
for which $\pi(p)$, respectively $\pi_1(p)$, is small. As we know from the previous section, in the condensed phase
$\pi_1(p)$ is many orders of magnitude greater for large $p$ than for
small $p$. Therefore when $m$ is of order $M$, and $n$ is of order $1$, 
the time $\tau_{mn}$ varies very slowly with $m$ and, on the other hand, it is
very sensitive to $n$. We thus put $m=M$ for simplicity and concentrate on
$\tau_{Mn}$.


\subsection{Star graph}
We assume $u(p)=1$ as before. 
Inserting the expression (\ref{pi1star}) for the particle occupation distribution $\pi_1(m)$ for the central node 
of the star into Eq.~(\ref{Tmngen}) in place of $\pi(m)$ we obtain
\bq
\tau_{mn} = \sum_{p=n+1}^{m} \sum_{l=p}^{M} (N-1)^{l-p} 
\frac{(M+N-l-2)!(M-p)!}{(M+N-p-2)!(M-l)!} .
\label{tmnstar}
\eq
From Sec.~II~B we know that the condensate contains  
$m =\left<m_1\right> \approx M$ balls in the steady state and that fluctuations 
are very small. This justifies the choice $m=M$ we made above.
Changing the summation variables similarly as in the previous section
we find:
\ba
\tau_{Mn} = (N-2)! \sum_{r=0}^{M-n-1} \frac{r!}{(N-2+r)!} (N-1)^r \times \nonumber \\
	\times \sum_{q=0}^r (N-1)^{-q} \frac{(N-2+q)!}{q!(N-2)!}.
\ea
In the second sum we can move the upper limit to infinity using exactly
the same approximation as in Sec.~II~B:
\bq
\tau_{Mn} \approx \left(\frac{N-1}{N-2}\right)^{N-1}(N-2)!  
\sum_{r=0}^{M-n-1} \frac{r!(N-1)^r}{(N-2+r)!} .
\eq
After a variable change $r\to M-n-1-r$, the remaining sum can be approximated as
\ba
	\sum_{r=0}^{M-n-1} \frac{(M-n-1-r)!}{(M+N-n-3-r)!} (N-1)^{-r} \approx  \nonumber \\
	\approx \frac{(M-n-1)!}{(M+N-n-3)!}	 \sum_{r=0}^\infty \left(\frac{M+N-n-3}{(M-n-1)(N-1)}\right)^r, \hspace{-5mm} \nonumber \\
	\label{eq:sumr0}
\ea
such that we finally arrive at the formula:
\ba
\tau_{Mn} \approx \left(\frac{N-1}{N-2}\right)^{N}(N-2)! (N-1)^{M-n-1} \times \nonumber \\
\times \frac{M-n-1}{M-n-2} \frac{(M-n-1)!}{(M+N-n-3)!}.
\label{TMnstar}
\ea
We see that the presence of $(N-1)^{-n}$ makes the time $\tau_{Mn}$
indeed very sensitive to $n$.
In Fig.~\ref{f4} we see $\tau_{M0}$ compared to computer simulations.
This complicated formula has a simple behavior in the limit of
very large systems and for $n=0$. In the limit of large $M$ and 
for $N$ being fixed, the time $\tau_{Mn}$ grows exponentially with $M$,
\bq
\tau_{M0} \sim (N-1)^M = e^{\mu M},	\label{tm0ap1}
\eq
with $\mu=\log(k_1/k)=\log(N-1)$, while for fixed density $\rho=M/N$ 
and $N\to\infty$ it increases faster than exponentially,
\bq
\tau_{M0} \sim e^{\rho N \log N}. \label{tm0ap2}
\eq
The approximate formulas (\ref{tm0ap1}) and (\ref{tm0ap2}) 
can be alternatively obtained using a kind of Arrhenius law \cite{arh,god},
which states that the average life time is inversely proportional to the 
minimal value of the occupation number distribution:
\bq
\tau_{m0} \sim 1/\pi_1({\rm min}),
\eq
where one thinks about the condensate's melting as of tunneling 
through the potential barrier in a potential $V(m)=-\log \pi_1(m)$. In our case the potential $V(m)$ grows monotonically with $m$ going to zero, so the ball rather bounces from the wall at $m=0$ than tunnels through it, but the reasoning is the same.
From Eq.~(\ref{pi1star}) we have $\pi_1({\rm min})\sim (N-1)^{-M}$ for 
fixed $N$ and large $M$ and we thus get again Eq.~(\ref{tm0ap1}), while 
for fixed density $\pi_1({\rm min})$ falls over-exponentially 
which results in Eq.~(\ref{tm0ap2}).
\begin{figure}
\psfrag{x}{$M$}
\psfrag{y}{$\tau_{M0}$}
\includegraphics[width=8cm]{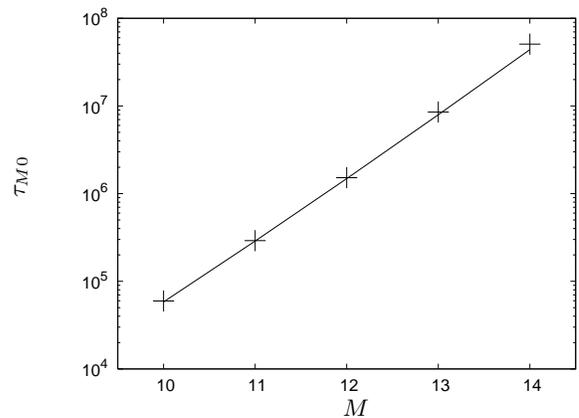}
\caption{The average lifetime $\tau_{M0}$ for the star graph with 
$N=10$ calculated from Eq.~(\ref{TMnstar}) (solid line) and found in 
computer simulations (points).}
\label{f4}
\end{figure}

So far we have discussed the singular node. It is quite surprising that
the formula (\ref{Tmngen}) works also well for regular nodes.
 If we blindly substitute $\pi_1(k)$ by $\pi_i(k)$ we get the following expression for $\tau_{i,mn}$:
\ba
& &\tau_{i,mn} = \nonumber \\
& &\frac{(N-1)^n-(N-1)^m+(m-n)(N-2)(N-1)^M}{(N-2)^2(N-1)^{M-1}}. \nonumber \\
\ea
For $m$ fixed, the transition time decreases 
almost linearly with $n$. The typical occupation of the regular node
is much smaller than $M$, so we concentrate on $m\ll M,n=0$ and $N\gg 1$.
The approximate formula reads:
\bq
\tau_{i,m0} \approx m \frac{N-1}{N-2}	\label{tm0ext},
\eq
and grows very slowly in comparison to the life time of the condensate 
at the singular node. This linear growth can be easily understood as 
the minimal time needed for $m$ particles to hop out from a regular node.
One must remember that in practice we cannot observe transitions for large $m$, because the 
probability of having such states is extremely small as it stems from 
Eq.~(\ref{pistext}) for $\pi_i(m)$.

\subsection{Single inhomogeneity}
Next we consider the single inhomogeneity graph from Sec.~II~C. 
In the condensed phase the occupation $m$ fluctuates quickly 
around $\left<m_1\right>$ and even if it is smaller than $M$ we can
assume that $\tau_{mn}\approx \tau_{Mn}$ because the transition time $\tau_{Mm}$
from $M$ to $m$ balls is very small in comparison to $\tau_{Mn}$ .
Therefore we shall concentrate again on $\tau_{Mn}$. 
From Eqs.~(\ref{pi1inh}) and (\ref{Tmngen}) we have
\bq
\tau_{Mn} = \sum_{p=n+1}^M \sum_{l=p}^M \alpha^{p-l} 
\frac{\binom{M+N-l-2}{M-l}}{\binom{M+N-p-2}{M-p}}.
\eq
Changing variables we get
\ba
\tau_{Mn} = \sum_{p=n+1}^M \frac{(M-p)!}{(M+N-p-2)!} \times \nonumber \\
\times \sum_{q=0}^{M-p}\alpha^{-q} \frac{(M+N-p-q-2)!}{(M-p-q)!}.
\ea
The sum over $q$ can be approximated by an
integral which can then be estimated by the saddle-point method.
The saddle point is $q_*=\alpha (N-2)/(1-\alpha)$ as in 
Sec.~II~C and therefore all calculations are almost identical. In 
this way we obtain
\bq
\sum_q \dots \approx \alpha^p \times \alpha^{\alpha\frac{N-2}{1-\alpha}-M}
\frac{\left(\frac{N-2}{1-\alpha}\right)!}{\left(\frac{\alpha(N-2)}{1-\alpha}\right)!}
\sqrt{\frac{2\pi \alpha (N-2)}{(1-\alpha)^2}},
\label{factor}
\eq
where we used again the notation 
$x!\equiv \Gamma(1+x)$.
The only dependence on $p$ in this expression is through the factor $\alpha^p$.
Thus to calculate $\tau_{Mn}$ it suffices to evaluate the sum
\bq
  \sum_{p=n+1}^M \alpha^p \frac{(M-p)!}{(M+N-p-2)!}. \label{eq:sumpn}
\eq
Because every term in the sum is proportional to $1/\pi_1(p)$ from Eq.~(\ref{pi1inh}), in the condensed phase the function under the sum has a minimum at the saddle point $p_*=m_*$. As this minimum is very deep, the effective contribution to the sum can be split into two terms: for small $p\ll m_*$ and for $p\gg m_*$. The ``small-$p$'' part can be evaluated like in Eq.~(\ref{eq:sumr0}) by pushing the upper limit to infinity and approximating the ratio of factorials by some number to power $p$. To calculate the ``large-$p$'' part it is sufficient to take the last two terms in Eq.~(\ref{eq:sumpn}), namely for $p=M$ and $p=M-1$, because they decrease quickly. The complete formula for $\tau_{Mn}$ is finally given by
\ba
	\tau_{Mn} \approx \alpha^{\alpha\frac{N-2}{1-\alpha}-M} \frac{\left(\frac{N-2}{1-\alpha}\right)!}{\left(\frac{\alpha(N-2)}{1-\alpha}\right)!}
	\sqrt{\frac{2\pi \alpha (N-2)}{(1-\alpha)^2}} \times \nonumber \\ 
	\times \left[ \frac{M!}{(M+N-2)!}  \left(\alpha\frac{M+N-2}{M}\right)^{n+1} \times \right. \nonumber \\
	\left. \times \left( 1-\alpha\frac{M+N-2}{M}\right)^{-1} +  \frac{\alpha^{M-1}(\alpha(N-1)+1)}{(N-1)!} \right]. \nonumber \\
	\label{compl}
\ea
In Fig.~\ref{f5} we compare this theoretical formula with $\tau_{M0}$ from numerical simulations. 
Equation (\ref{compl})
simplifies in the limit of large systems. When 
one allows $M\to\infty$ while keeping $N$ and $\alpha$ fixed,
then the life time grows exponentially,
\bq
\tau_{M0} \sim \left(\frac{1}{\alpha}\right)^M = e^{\mu M}. \label{tm0inh}
\eq
For $\mu=\log(N-1)$, that is for a star graph, it reduces to the formula (\ref{tm0ap1}).
In the limit of fixed density $\rho=M/N>\rho_c$ 
and for $N,M\to\infty$:
\bq
\tau_{M0} \sim e^{N\left[ -\log(1-\alpha)+
\rho\log(\rho/\alpha)-(1+\rho)\log(1+\rho)\right]}. \label{last}
\eq
We see that the life time grows exponentially only if \mbox{$k_1>k$}, that is for positive external field $\mu=\log(k_1/k)$.
As before, we can explore the limit when the single inhomogeneity graph reduces to a star graph. 
Inserting $\mu=-\log\alpha=\log(N-1)$ into Eq.~(\ref{last}) we recover Eq.~(\ref{tm0ap2}) as the leading term
for large $N$.
\begin{figure}
\psfrag{x}{$M$}
\psfrag{y}{$\tau_{M0}$}
\psfrag{y2}{\hspace{-5mm}$\log \tau_{M0}$}

\includegraphics[width=8cm]{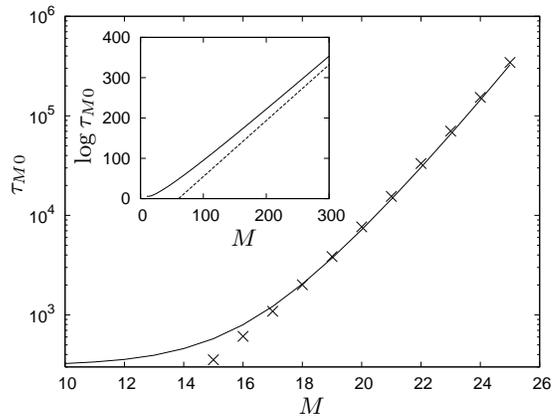}
\caption{Comparison between the ``experimental'' results (points) and the
theoretical prediction (\ref{compl}) (solid line) for $\tau_{M0}$ of a $4$-regular graph
with single inhomogeneity $k_1=16$. The graph size is $N=20$. The inset compares $\tau_{M0}$ calculated from Eq.~(\ref{compl})
(solid line) with that from Eq.~(\ref{tm0inh}) (dashed line), valid in the large-$M$ limit. The prefactor
in Eq.~(\ref{tm0inh}) is chosen to match the $M\to\infty$ limit of the two formulas.}
\label{f5}
\end{figure}

\section{Conclusions}
In this paper we have studied static and dynamical properties of the condensation
in zero-range processes on inhomogeneous networks. We have focused on the case where the network
is almost a $k$-regular graph except that it has a single node of degree $k_1$
larger than $k$. This type of network could be a rude prototype 
of inhomogeneities encountered in scale-free networks having
a single hub with very high degree and many nodes of much smaller degrees.
Indeed, from the point of view of the hub, 
the remaining nodes look as if they formed an almost homogeneous system.
We have shown that the distribution of balls $\pi_1(m)$ at the singular node has a maximum at $m\approx N(\rho-\rho_c)$ where $\rho_c$ is the critical density above which the condensate is formed. The average occupation of regular nodes is equal to $\rho_c$ and the condensate never appears on them. However, the condensate is not a static phenomenon. It fluctuates and it sometimes
melts and disappears from the singular node. Then the surplus of balls distributes uniformly on all other nodes. After a while the condensate reappears and its typical life time $\tau$ grows exponentially like $e^{\mu M}$, where $M$ is the number of balls and $\mu=\log (k_1/k)$ plays the role of an external field explicitly breaking the permutational symmetry of the system. This behavior is qualitatively distinct from that observed in homogeneous systems with a power-law distribution of balls, where $\tau$ grows only like a power of $M$, and the symmetry is spontaneously broken. Thus the transition $\mu=0\to \mu> 0$ changes dramatically all properties of the system.

In all above calculations we assumed for simplicity that the hop rate $u(m)=1$ and thus for the homogeneous system there would be no condensation. However, in the case $u(m)=1+b/m$ which produces the power-law in $\pi(m)$ for regular graphs, in an inhomogeneous system, apart from the condensation on the singular node, we would expect a second condensate on some regular node if the number of particles would be large enough to exceed the critical density for the homogeneous sub-system. Thus we could expect the presence of two critical densities $\rho_{1c}$ and $\rho_{2c}$ and two condensates having completely distinct properties. We leave this interesting question for future research.

\section{Acknowledgments}
We acknowledge support by the EC-RTN Network ``ENRAGE'' under 
grant No. MRTN-CT-2004-005616, an Institute Partnership grant 
of the Alexander von Humboldt Foundation,
a Marie Curie Host Development Fellowship under 
Grant No. IHP-HPMD-CT-2001-00108 (L.B. and W.J.),  
a Marie Curie Actions Transfer of Knowledge project ``COCOS'',
Grant No. MTKD-CT-2004-517186 and a Polish Ministry of Science
and Information Society Technologies Grant 1P03B-04029 (2005-2008)
(Z.B.). B.W. thanks the German Academic Exchange Service (DAAD) for a fellowship.

\end{document}